\documentclass[12pt,preprint,a4paper]{aastex}
\usepackage{times}
\usepackage{graphics,epsf}
\usepackage{amsmath}                
\usepackage{amsfonts}               
\usepackage{amssymb}                
\usepackage{epsfig}                 
\usepackage{graphicx}
\usepackage{url}
\usepackage{tabularx}
\usepackage{booktabs}

\def \km{~\rm{km}}

\begin{document}

\title{A formation scenario for the triple pulsar PSR J0337+1715: breaking
 a binary system inside a common envelope}

\author{Efrat Sabach \altaffilmark{1} and Noam Soker\altaffilmark{1}}

\altaffiltext{1}{Department of Physics, Technion -- Israel
Institute of Technology, Haifa 32000, Israel;
efrats@physics.technion.ac.il; soker@physics.technion.ac.il}

\begin{abstract}
We propose a scenario for the formation of the pulsar with two
white dwarfs (WDs) triple system PSR~J0337+1715.
In our scenario a close binary system is tidally and frictionally
destroyed inside the envelope of a massive star that later goes
through an accretion induced collapse (AIC) and forms the neutron star (NS).
The proposed scenario includes a new ingredient of a binary system
that breaks-up inside a common envelope.
We use the {\it binary\_c} software
to calculate the post break-up evolution of the system,
and show that both low mass stars end as helium WDs.
One of the two lower mass stars that ends further out, the tertiary star,
transfers mass to the ONeMg WD remnant of the massive star, and
triggers the AIC.
The inner low mass main sequence star evolves later, induces AIC if the
tertiary had not done it already, and spins-up the NS to form a millisecond pulsar.
This scenario is not extremely sensitive to many of the parameters, such as the
eccentricity of the tertiary star and the orbital separation of the secondary
star after the low mass binary system breaks loose inside the envelope, 
and to the initial masses of these stars. 
The proposed scenario employs an efficient envelope removal by jets
launched by the compact object immersed in the giant envelope, and
the newly proposed grazing envelope evolution.
\end{abstract}

\keywords{
binaries: close --- pulsars: individual (PSR J0337+1715) --- stars:
 mass-loss --- stars: neutron}

\section{INTRODUCTION}
\label{sec:intro}

Despite much progress, e.g. review by \cite{Ivanovaetal2013},
the common envelope (CE) phase is one of the least understood
processes in stellar binary evolution.
The evolution becomes much more complicated if a close tertiary
body exists in the system, such that the CE evolution includes
three gravitating bodies.
The evolution that formed the Galactic millisecond pulsar (MSP) triple
system PSR~J0337+1715 might have gone through a CE phase involving
three stars.

PSR~J0337+1715 is a triple system that contains a $1.438 M_\sun$
radio MSP with a spin period of $P = 2.73$~ms
orbited by two white dwarfs (WDs) \citep{Ransometal2014}.
The inner WD mass, orbital period, and eccentricity are
$M_{\rm WD,2}=0.197 M_\sun$, $P_{\rm 1,2}=1.63$~day, and
$e_{1,2}= 6.9 \times 10^{-4}$, respectively, while those of the
outer WD, hereafter the tertiary star, are
$M_{\rm WD,3}=0.410 M_\sun$, $P_{\rm 12,3}=327$~day, and $e_{12,3}=0.035$.
The system is coplanar to within $0.01 ^\circ$. Deeper study of
the inner WD was conducted by \cite{Kaplanetal2014}.

\cite{TaurisHeuvel2014} propose that the progenitor of
PSR~J0337+1715 contained a massive main sequence (MS) star of mass
$\approx 10 M_\sun$, orbited by two lower mass MS stars having
masses and orbital periods of $1.1 M_\sun$ and 835~day, for the
inner star, and $1.3 M_\sun$ and 4020~day for the outer stars. One
after the other the two stars entered the bloated envelope of the
massive star as it became a giant, and ended at very short
orbital periods. The neutron star (NS) was formed by a core
collapse supernova (CCSN) of the primary star.
\cite{TaurisHeuvel2014} find the range of parameters that can work
for their scenario to be about $20 \%$.
Later we find that for our proposed scenario there is a similar freedom in
the parameter space that can work.
However, we see two problematic steps in this evolution.
(1) It is not clear how the two stars ended at short orbital periods.
If the inner star managed to eject the envelope, as is assumed in the CE
evolution, then there will be no envelope left to bring the outer tertiary
star to a short orbital period.
If on the other hand there is enough mass left in the primary
envelope to swallow the tertiary star and bring it closer, then
we expect the friction to cause the secondary star to collide with
the primary core.
We avoid this problem by taking the two outer stars to enter the
giant's envelope at the same time as a tight binary system.
(2) NSs that are born in CCSNe are born with a non-zero velocity, called a
kick velocity.
As it is very unlikely that the system will stay
coplanar after a kick, in the scenario of \cite{TaurisHeuvel2014}
the system gains coplanar geometry at a later phase, when
the tertiary (outer) star overflows its Roche lobe and transfers
some mass to the inner binary system, now composed of a NS and a
MS star. It is still needed to be shown that such a
process is capable in bringing a triple system to coplanar
geometry.
In our proposed scenario the NS is formed through accretion
induced collapse (AIC) of an ONeMg WD
\citep{Nomotoetal1979, Canaletal1980, Nomoto&Kondo1991}.
The amount of baryonic mass ejected in this process is very small, while
the neutrinos are expected to escape in a spherical geometry.
No kick perpendicular to the orbital plane is expected.
The only kick is a small one in the orbital plane due to the orbital motion
of the WD around the center of mass of the triple system.

In the standard model MSPs are formed in low mass X-ray binaries
(LMXBs) and spin up due to accretion of mass and angular
momentum in a Roche lobe overflow (RLOF) from a companion star
(\citealt{Bhattacharya&vandenHeuvel1991, Tauris&vandenHeuvel2006, Tauris2015},
and references therein).
In our scenario most likely the tertiary star causes the AIC
of the primary star that forms a NS, and the secondary star transfers
mass to the primary star and causes it to spin up and become a MSP.
Triple star evolution has been proposed for binary pulsars
(e.g. \citealt{Freireetal2011, PortegiesZwartetal2011, Pijlooetal2012}),
such as the peculiar system PSR~J1903+0327 \citep{Campionetal2008}.
\cite{Freireetal2011} proposed that two MS stars were spiralling
together in the envelope of the primary massive progenitor of the
pulsar PSR~J1903+0327.
The inner companion was later destroyed or ejected, leaving behind the
binary pulsar.
Again, it is not clear that one envelope can cause substantial
spiralling-in of two well separated stars (the secondary and the tertiary).

The problems we find in the scenario proposed by
\cite{TaurisHeuvel2014} motivate us to consider a different
scenario, also involving a CE phase with three stars, for the
formation of PSR~J0337+1715. The large fraction of triple systems
among binary stars, e.g., \cite{Rappaportetal2013} who argue that
at least $20 \%$ of all close binaries have tertiary companions,
further motivates us to consider different types of triple systems
than previously proposed models for binary and triple pulsar
systems. In section \ref{sec:scenario} we describe our proposed
scenario.
We employ some well studied processes in binary pulsars,
e.g., AIC, evolution of close LMXBs, termination of
the secondary evolution when its core mass is $\approx 0.2M_\odot$,
and more \citep{Refsdal&Weigert1971, Webbinketal1983, Savonije1987,
Tauris&Savonije1999, Taurisetal2013, Ablimit&Li2015, Tauris2015}.
Nevertheless, we also include new ingredients. The new main ingredient,
breaking-up a close
binary system inside a CE, is presented in section
\ref{sec:binary_CE}. In section \ref{sec:pulsar_formation} we
study the evolution of the secondary and tertiary stars to form
WDs. Our discussion and summary are in section \ref{sec:Summary}.

\section{THE SCENARIO}
\label{sec:scenario}
\begin{figure}[h!]
\centering
\includegraphics[width=150mm]{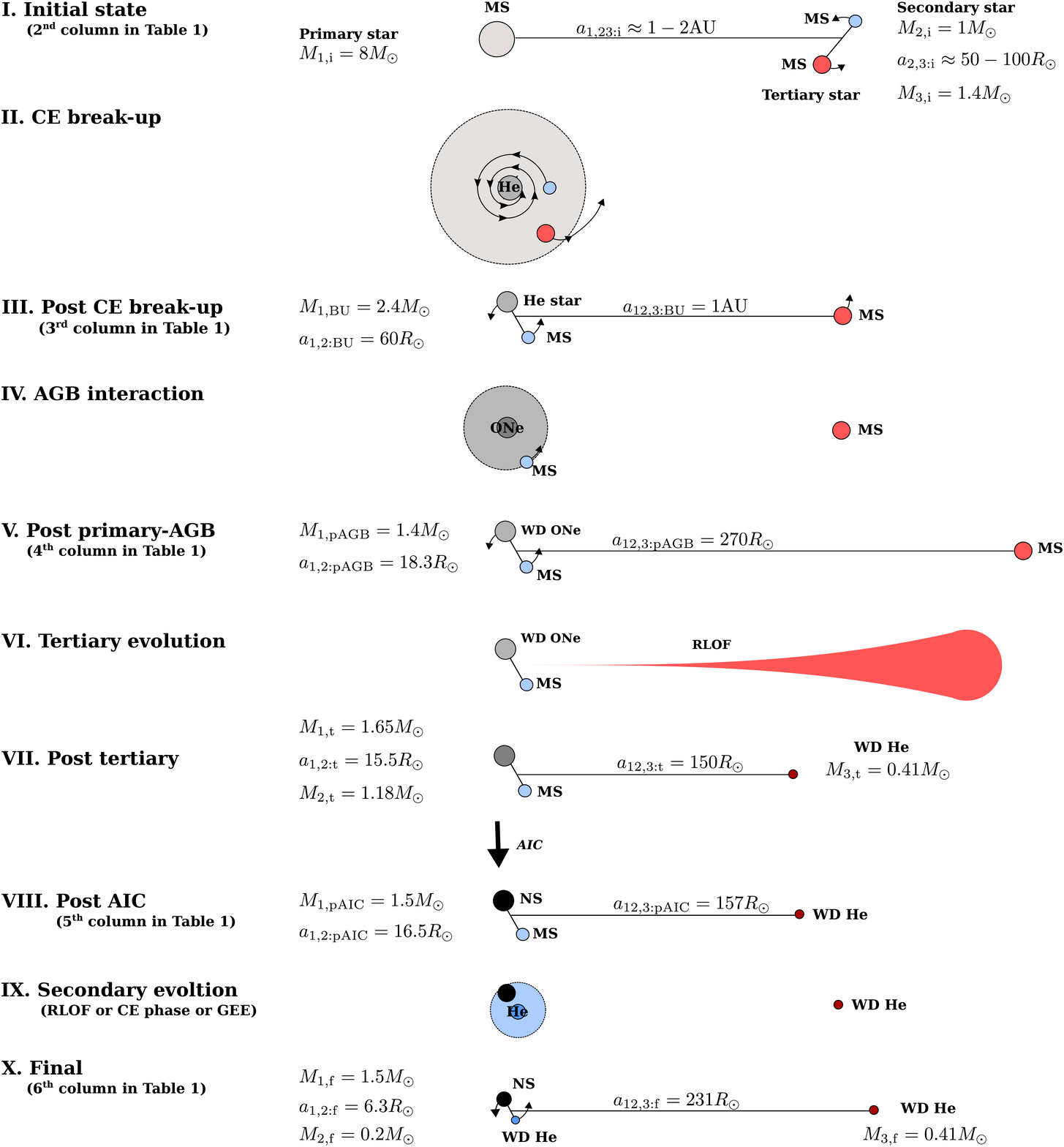}
\caption{The evolution of the triple system PSR~J0337+1715 from the
initial setup, where all three stars are on the main sequence (MS), and
until the current observations.
The initial masses and orbital separations represent a case study.
There is a range of parameters for which the scenario can work.
Note that in our notation the initial mass of the tertiary star is
larger than that of the secondary star: $M_{1,i}>M_{3,i}>M_{2,i}$.
The colors are as follows: grey is for the primary star, blue for the
secondary star and red for the tertiary star. As each star reaches a
more advanced evolutionary stage its color darkens.
The evolution from phases I to III is assumed here.
The evolution there after (phases III - X) is calculated with
{\it binary\_c}.
Stage IX can occur in one of three processes: stable RLOF
with some mass loss from the second Lagrangian point, an efficient CE
phase, or a GEE.
We here present the CE possibility since it is the obtained secondary
evolution in {\it binary\_c}.
}
\label{fig:mech}
\end{figure}

The proposed scenario is described schematically in Fig.\ref{fig:mech}
and parameters for one case studied here with {\it
binary\_c} (\citealt{Izzardetal2004, Izzardetal2006,
Izzardetal2009}; section \ref{sec:pulsar_formation}) are listed in
Table \ref{tab:parameters}.
The initial triple-star system is composed from a primary star,
the initially more massive MS star, and a tight binary
system of lower mass MS stars.
The outer binary system is tight in the sense that the orbital
separation of the two outer stars  is much smaller than the
orbital separation between their center of mass and the massive star.
The interaction starts when the primary star becomes a giant, and its
envelope tidally interacts with the tight binary system.

We divide the evolution of the system into five main stages
separated by four phases of strong stellar interaction.
Here we describe the four interaction phases.
The first one is further
discussed in section \ref{sec:binary_CE}, and the last three
interaction phases are studied in section
\ref{sec:pulsar_formation}.

(1) {\it Tidal break-up inside a common envelope.} The first stage is
the most delicate one, as we suggest that two processes occur
simultaneously:
The chaotic process of tidal break-up and the poorly understood
CE evolution.
We start (phase I in Fig.\ref{fig:mech}) with a triple system
composed of three MS stars of masses
$M_{\rm 1:i} \approx 7.5-8.5 M_\odot$, $M_{\rm 2:i} \approx 0.9-1.2M_\odot$,
and $M_{\rm 3:i} \approx 1.3-1.4$, which we term the primary, 
secondary and tertiary, respectively.
The two lighter stars form a tight binary system, for which we estimate
the initial orbital separation for a circular (or low eccentricity) orbit
to be $a_{\rm 2,3:i} \approx 50-100 R_\odot$.
This estimate will have to be checked with a three-body numerical code
that includes tidal interaction and mass removal by jets from the two stars.
This is way beyond the scope of this paper.   
The primary evolves and suffers two expansion episodes, when it has
a helium core and later when the core is made up mainly from CO.
Because of tidal interaction, most likely the CE occurs in the first
expansion episode (phase II in Fig.\ref{fig:mech}).
The outer binary system enters the envelope of the giant,
spirals-in, and removes the hydrogen-rich envelope.

As the outer binary spirals in, it breaks apart because of tidal
interaction.
We estimate this to occur when the center of mass of the
outer binary is at  $a_{\rm 1,23}\approx a_{\rm 2,3}\approx  100R_\odot$
from the center of the primary.
The secondary star, the lighter one, loses angular momentum more
efficiently and spirals-in to an orbital separation of $\approx 50-70R_\odot$
from the $ 2.4M_\sun$ helium core of the primary, now having a radius of a
${\rm few}\times R_\odot$.
The tertiary star ends further out.
This is further discussed in section \ref{sec:binary_CE}.

(2) {\it AGB interaction.}
As the primary continues to evolve and forms an ONe core (phase
IV in Fig.\ref{fig:mech}), the secondary is as massive as the
primary's envelope and is able to bring it to co-rotation, enhance
the mass loss rate \citep{ToutEggleton1988} and expel the entire
envelope of the primary (phase V in Fig.\ref{fig:mech}).
Moreover, it is possible that the secondary accretes mass through
an accretion disc and launches two opposite jets that expel part
of the primary He-rich envelope \citep{Soker2014, Soker2015}.
If a CE prescription is used, like what we will do later, then the
removal of an envelope by spin-up and jets leads to an effective
value of the $\alpha_{\rm CE}$ parameter that is larger than unity,
$\alpha_{\rm CE-eff} > 1$.
At the end of this phase of evolution the primary forms a
$1.25-1.4M_\sun$ ONeMg WD,
and the triple system has the properties listed in the fourth
column of Table \ref{tab:parameters} for a case study.
These parameters are justified in the next two sections.
In the table, circular orbits are assumed
for the initial system.
\begin{table}[htbp]
\footnotesize
    \centering
    \caption{Stellar evolution of a case study}
    \centering                                              
\scalebox{0.95}{
    \begin{tabular}{c c c c c c c}                                    
    \hline                                                  
    \hline                                                  
    \multicolumn{1}{c} -     & Initial			         & post break-up       & post primary-AGB        & post AIC		& final stage 	   & Observed \\
   \\[0.1ex]
    \hline                          
    $M_1 (M_\sun) $         &$8$ (MS)                     &$2.4$ (He star)      &$1.4$ (ONeMg WD)           &$1.5$ (NS)          &$1.5$ (NS)         &$1.4378$ \\
    $M_2 (M_\sun)$          &$1$   (MS)                   &$1$ (MS)           &$1$ (MS)               &$1.18$ (MS)         &$0.2$ (He WD)      &$0.1975$ \\
    $M_3 (M_\sun)$          &$1.4$ (MS)                   &$1.4$ (MS)           &$1.4$ (MS)               &$0.41$ (He WD)      &$0.41$ (He WD)     &$0.4101$ \\
    $a_{\rm inner} [R_\sun]$&$a_{\rm 2,3:i}\approx 50-100$&$a_{\rm 1,2:BU}=60$  &$a_{\rm 1,2:pAGB}=18.3$  &$a_{\rm 1,2:pAIC}=16.5$&$a_{\rm 1,2:f}=6.3$&$6.83$   \\
    $a_{\rm outer} [R_\sun]$&$a_{\rm 1,23:i}\approx 200-400$&$a_{\rm 12,3:BU}=214$&$a_{\rm 12,3:pAGB}=270$&$a_{\rm 12,3:pAIC}=157$&$a_{\rm 12,3:f}=231$&$252.5$  \\
    Comments                &$e_{\rm 2,3:i}=0$            &$e_{\rm 1,2:BU}=0.5$  &$e_{\rm 1,2:pAGB}=0$    &$e_{\rm 12,3:pAIC}=0$     \\
                            &$e_{\rm 1,23:i}=0$           &$e_{\rm 12,3:BU}=0.5$ &                         \\
 \\

    \hline                                          
\end{tabular}
}
\footnotesize \flushleft The masses and orbital separations of
the triple system PSR~J0337+1715 at the main phases of the evolution, for a
case study calculated with {\it binary\_c} (Izzard et al 2004, 2006, 2009).
Circular orbits are assumed for the initial setup.
The table shows an example of one case study with certain initial
parameters (column 2; phase I in Fig.\ref{fig:mech}) out of the allowed
parameter range given in the text, e.g., the primary initial mass
$M_{\rm 1:i}\approx 7.5-8.5 M_\odot$, secondary initial mass
$M_{\rm 2:i} \approx 0.9-1.2M_\odot$, and tertiary initial mass
$M_{\rm 3:i}\approx 1.3-1.4$. 
The post break-up (BU) properties are listed in column 3
(phase III in Fig.\ref{fig:mech}).
Note the exchange of inner and outer binary stars from the initial stage to
the post break-up stage: the initial tight binary is composed of the
secondary and tertiary stars, whereas after the break-up of the secondary-
tertiary binary system inside the common envelope, the primary and secondary
stars become the inner binary system.
The post primary-asymptotic giant branch (pAGB) properties are listed
in column 4 (phase V in Fig.\ref{fig:mech}), where we find the allowed
primary post-AGB mass to range over $M_{\rm 1:pAGB}\approx 1.25-1.4 M_\odot$.
During the following evolution phase of the tertiary (phase VI in
Fig.\ref{fig:mech}) we find with {\it binary\_c} that the tertiary transfers
a mass of $M_{\rm acc:t} = 0.55 M_\odot$ via RLOF to the inner binary.
We take $0.44 M_\odot$ to be accreted by the central binary system
taking into account mass loss by jets in the accretion disc around
the inner binary.
Consequently, the primary experiences an AIC and forms a NS.
We assume its mass is reduced due to energy carried by neutrinos emitted
by the cooling NS.
The post AIC parameters are listed in column 5 (phase VIII in Fig.\ref{fig:mech}).
The next stage is the evolution of the secondary star,
(phase IX in Fig.\ref{fig:mech}) after which the system reaches its final
stage (column 6; phase X in Fig.\ref{fig:mech}) that corresponds with the
observed parameters of
the system (column 7).
 \label{tab:parameters}
\end{table}

(3) {\it Tertiary evolution.} As the tertiary evolves and becomes
an evolved red giant branch (RGB) star it interacts with the inner
binary system, now composed of an ONeMg WD and a MS star 
(phase VI in Fig.\ref{fig:mech}).
Two processes take place and lead to the next stage of the triple
system that we term {\it post tertiary}.
The first is tidal interaction that leads to the circularization
of the orbit, even if the orbit is very eccentric.
For an eccentricity of $e_{\rm 12,3:pAGB}=0.8$ we find the required
semimajor axis before the evolution of the tertiary (the post
primary-AGB stage) to be $a_{\rm 1,23:pAGB} \approx 550 R_\sun$.
This implies a periastron distance of $\approx 110 R_\sun$ which
leads to a very strong tidal interaction near periastron passages
and subsequently to a circular orbit.
These parameters will bring the final result of our simulation even
closer to the observed parameters of the system.

The second process is mass transfer.
We find that mass transfer might cause the ONeMg WD to
overpass the Chandrasekhar mass limit and suffer an AIC
(e.g., \citealt{Taurisetal2013, Ablimit&Li2015}, for recent papers on the
formation of MSPs by AIC).
Although there are large uncertainties, from {\it binary\_c} we find
that the duration of the RLOF from the tertiary star might allow the
AIC of $M_1$.
The tertiary star transfers $\approx 0.5M_\odot$ to the central binary system
at an average rate of $\approx 1.8\times 10^{-7}M_\odot \rm yr^{-1}$.
This will result in an average mass transfer rate of
$\approx 1\times 10^{-7}M_\odot \rm yr^{-1}$ on to $M_1$.
This mass transfer is about the rate required to cause an AIC
\citep{Taurisetal2013, Ablimit&Li2015}.
\cite{Taurisetal2013} found that for the case of a giant donor star there is
a narrow region of donor star masses and initial orbital periods that
allows AIC: $0.9\lesssim M_{\rm donor}\lesssim 1.1 M_\odot$ and
$100 \rm d \lesssim  P_{\rm orb}\lesssim 300 \rm d$.
Though $M_2$ is a bit larger than this upper mass limit for $M_{\rm donor}$ the
initial period and the estimated mass transfer rate found here are in
the range required for an AIC.
The tertiary mass can be fitted to the observed value, as we show in
section \ref{sec:pulsar_formation}.
The mass transfer increases also the mass of the secondary star and
leads to a shrinkage in the orbit of the inner binary system
(phase VII in Fig.\ref{fig:mech}).
In case that the mass transfer from the tertiary star is not sufficient
for the primary to undergo AIC, only a small amount of mass will accumulate
on $M_1$.
The subsequent evolution of the secondary star, $M_2$, will lead the
ONeMg WD, $M_1$, to experience AIC and become a NS.
We do not elaborate on this possibility since we find the AIC via the
tertiary star to be more likely.

We take the resulting NS from the AIC to have zero kick velocity as
is generally agreed upon from simulations of an AIC event
\citep{Dessartetal2006,Kitauraetal2006}.
Mass loss by neutrinos can result in a small kick velocity in the orbital
plane of the system due to the orbital motion of the WD around the
system's center of mass.
The NS symmetrically loses $\approx0.15M_\odot$ by neutrinos
($1/20$ of the total mass of the system), with the local NS velocity of 
$\approx 120 \km\; \sec^{-1}$ (taking the system parameters at phase VII
in Fig.\ref{fig:mech}).
Due to momentum conservation this will result in a small recoil
of the entire system ($\approx 6 \km\; \sec^{-1}$) in the plane of
the system alone, without deviating from a coplanar geometry.

(4) {\it Secondary evolution.} The observed mass of the secondary
star is $0.1975 M_\sun$, and it is a He WD.
This fits the core of a low mass star during its post-MS evolution
\citep{Kippenhahn&Weigert1990}, as has been suggested before for low mass
binaries (\citealt{Refsdal&Weigert1971, Savonije1987}; 
See section \ref{sec:pulsar_formation} for a discussion on the
evolution of the remnant He WD).
In our scenario the secondary and the NS strongly interact when
the secondary evolves as a giant (phase IX in Fig.\ref{fig:mech}).
According to {\it binary\_c} the observed orbital separation of 
$a_{\rm 1,2:obs}=6.83 R_\sun$ dictates the orbital separation prior
to the AIC phase to be $a_{\rm 1,2:pAIC}\approx 15.5-18.5 R_\sun$.
The mass transfer process will occur either through a CE phase (as given by
{\it binary\_c}) or by RLOF from the donor star, $M_2$, to the central NS, $M_1$.
If only mass transfer through $L_1$ is considered, the orbital separation
increases (e.g., \citealt{Taurisetal2013}).
However, the convective envelope and relatively rapid rotation of the
secondary star result in a magnetically active star that is most likely to
posses an extended corona with a wind.
As the secondary star is already filling its Roche lobe, the corona is now
extending to the second Lagrangian point, $L_2$, and a large fraction of
the mass lost from the extended corona is likely to do so from nearby $L_2$.
As the specific angular momentum of the mass lost through $L_2$ is
much larger than that of the binary system, by a factor $>2$, the $L_2$
outflow reduces the specific angular momentum of the inner binary system.
This is will lead to a shrinkage in the orbital separation \citep{Livioetal1979}.
From the recent study by \cite{Akashietal2015} we estimate for the present studied
system that by losing  $\approx 5\%$ of the total binary mass, $\approx0.1M_\odot$,
through $L_2$ the orbital separation of the inner binary decreases by a factor
of $\approx 2$.
This results in a final orbital separation of $a_{\rm 1,2:f}\approx6-8 R_\sun$.

On the other hand, by evolving the inner binary using {\it binary\_c} we
find that the mass loss from $M_2$ occurs via a CE phase, instead of stable
mass transfer via RLOF.
The CE phase might be caused by the magnetic activity of the secondary star
that causes loss of angular momentum from the system.
To account for the observed (final) orbital separation of the inner binary,
$a_{\rm 1,2:obs}=6.83 R_\sun$, the CE must be very efficient in
removing the envelope.
Jets blown by the NS can account for $\alpha_{\rm CE-eff}>1$.
The process might even be a grazing envelope evolution
(GEE; \citealt{Soker2015}; see next section) rather than a CE evolution,
which implies an even larger value of $\alpha_{\rm CE-eff}$ (although
this is not a real CE phase).
The tidal interaction and CE evolution explain the observed circular
orbit.
When we use {\it binary\_c} (section \ref{sec:pulsar_formation}),
we can justify taking $\alpha_{\rm CE-eff}>1$.
During this phase the secondary loses mass, a small fraction of which
is accreted by the NS, the orbital separation of the inner binary
decreases and that of the tertiary increases.
Overall we point out that the mass transfer can occur in one of the three
processes: stable RLOF with some mass loss from the second Lagrangian point, an
efficient CE phase, or a GEE .
The final masses and orbital separations are listed in Table
\ref{tab:parameters}.

\section{BINARY INSIDE A COMMON ENVELOPE}
\label{sec:binary_CE}

The tight binary system (the two lower mass MS stars) tidally
breaks-up approximately when the orbital separation of its center of
mass from the primary is (\citealt{Milleretal2005, Sesanaetal2009})
\begin{equation}
a_{1,23} \simeq \left( \frac{3M_{\rm1:i}(a_{1,23})}{M_{\rm 2,3:i}}
\right)^{1/3} a_{2,3}. \label{eq:tidalbu}
\end{equation}
The ratio of orbital periods at that epoch is
\begin{equation}
\frac{P_{1,23}}{P_{2,3}} \simeq  \left( \frac{a_{1,23}}{a_{2,3}}
\right)^{3/2} \left( \frac{M_{\rm1:i}(a_{1,23})+ M_{\rm 2,3:i}}{M_{\rm
2,3:i}} \right)^{-1/2} \simeq 3^{1/2}  \left( \frac{
M_{\rm1:i}(a_{1,23})}{M_{\rm1:i}(a_{1,23})+ M_{\rm 2,3:i}} \right)^{1/2}
\simeq 1.5,
\label{eq:orbitalp}
\end{equation}
where $M_{\rm 1,i}(a_{1,23})$ is the primary mass inner to the location
of the tight binary, $a_{1,23}$, and $M_{\rm 2,3:i}$ is the combined
initial mass of the secondary and tertiary stars.
From this condition it is clear that to reach tidal break-up the
tight binary must spiral-in such that the orbit $a_{1,23}$ shrinks
faster than the orbit of the tight binary system $a_{2,3}$.
We now discuss the condition for this to occur.

The tight binary system loses angular momentum of motion around the
primary star inside the envelope within a few dynamical times.
This implies that the tidal break up
occurs on a scale not much shorter than the spiralling-in time.
Both processes must be considered simultaneously.
We can safely say that the friction force of the two MS stars in the
envelope cannot be neglected.
If one of the two MS stars is ejected within the envelope it might
lose momentum and stay bound, even if the star is ejected with an
initial velocity above the escape velocity.
As we later show, even a very high eccentricity of the outer star's
orbit, $e \sim 0.8$, will lead to the desired outcome.
In a tidal-break up usually the lower mass companion of the binary is
ejected.
Inside the CE, however, dynamical friction with the envelope must be
considered.
The outcome might be that the lower mass companion of the tight 
binary is spiralling-in deeper and the heavier companion ends on a
larger orbit, particularly since the difference in mass between
the two MS stars is not large.

From preliminary simulations (Michaely, E., 2015, private
communication) it turns out that when the tight binary is immersed
in the undisturbed giant envelope the tight binary system becomes
even harder (its orbit shrinks), and no tidal break-up occurs. 
The gravitational drag causes both MS stars to lose energy, such that
the orbital separation between them becomes tighter faster than the
time on which their distance from the core of the giant star shrinks.
In such a case no tidal break-up occurs.
However, if roughly only the star closer to the giant's core suffers
from gravitational drag, tidal break-up can occur.
The way to substantially reduce the drag on the outer MS
of the tight binary system is to remove the envelope outside the
orbit of the tight binary system.
The tight binary system actually experiences a grazing envelope
evolution (GEE; \citealt{Soker2015}) rather than a true CE evolution.
In the GEE the binary system might be considered to evolve in a state
of ``just entering a CE phase''.
The companion star in a regular binary system, or the tight binary
system in the triple-system case, removes the envelope beyond its orbit by
launching jets \citep{Soker2014}.
The removal of the envelope outside the orbit prevents the formation
of a full CE phase.
This efficient envelope removal increases the effective value of the
CE parameter to values of $\alpha_{\rm CE} > 1$ (even though no real
CE takes place, but rather a GEE).

\section{FORMING THE PULSAR AND THE TWO WHITE DWARFS}
\label{sec:pulsar_formation}

We use the population nucleosynthesis code {\it binary\_c} of Izzard et al
(2004, 2006, 2009) based on the Binary Star Evolution (BSE) code of
\cite{Hurleyetal2002}.
As {\it binary\_c} is a stellar population nucleosynthesis framework,
we use it to calculate the properties of the system at stages beyond the tidal
break-up of the tight binary, i.e.
the mass of the each star, orbital separation, etc.

To treat the triple system for each phase of strong interaction we simulate
a binary system.
We give a representative case for the scenario, and later show that there is
no need here for heavy fine-tuning, although the parameter space is narrow.
At the post CE break-up phase of the system (phase III in Fig.\ref{fig:mech})
the primary is a He star of mass $M_{\rm 1,BU}=2.4M_\odot$ in an inner
binary with the secondary star of $M_{\rm 2,BU}= 1M_\odot$.
The primary continues to evolve and forms an ONe core in the mass range of
$M_{\rm 1:pAGB}\approx 1.25-1.4M_\odot$ as it expands along the asymptotic
giant branch (AGB; phase IV in Fig.\ref{fig:mech}).

Now starts another phase of strong binary interaction.
The secondary is massive enough to bring the primary envelope to co-rotation,
and hence enhances the mass loss rate from the primary, up to a factor
$\gtrsim 100$, e.g., \cite{ToutEggleton1988}.
Another process that can aid in removing the primary He-rich envelope is the
launching of jets by the MS companion \citep{Soker2014, Soker2015}.
This stage of interaction with tidally-enhanced mass loss rate and possible
jets can not be modelled correctly with the {\it binary\_c} code.
We can only mimic these processes by taking the $\alpha_{\rm CE}$ parameter
at this stage to be larger than unity.
We here take $\alpha_{\rm CE}=5$.
From examining other values we estimate that to reach the observed parameters
of the PSR J0337+1715 system, the inner binary separation at the post CE
break-up phase should be in the range of $\approx 50 - 70R_\odot$.
Once the entire envelope of the primary is ejected it forms an ONeMg WD, while  
the secondary star remains a MS star (phase V in Fig.\ref{fig:mech}).
We term this phase the post primary-AGB phase of the system and we estimate
the separation of the inner binary at this stage to be
$a_{\rm 1,2:pAGB}\approx 17-21 R_\odot$.

We consider a case study for an example of a possible evolution.
We take the immediately post primary-AGB system to be composed of an
ONeMg WD of mass $M_{\rm 1:pAGB}= 1.4 M_\odot$, and two MS stars with
masses of $M_{\rm 2:pAGB}=1.0 M_\odot$ and $M_{\rm 3:pAGB}=1.4 M_\odot$.
The orbital separation of the inner MS star and the ONeMg WD is
$a_{\rm 1,2:pAGB}= 18.3 R_\odot$, and that of the tertiary star from the
center of mass of the inner binary is $a_{\rm 3:pAGB}= 270 R_\odot$.
Circular orbits are assumed for the representative case.

We first follow the evolution of the outer MS star$-$the tertiary star, as
it is more massive than the secondary star.
For the usage of {\it binary\_c} we need to treat the inner binary
(the ONeMg WD + the low mass MS secondary star) as one star when following
the evolution of the outer MS star.
Numerically we take the inner binary system as a compact body of mass
$M_{\rm in-binary:pAGB}=M_{\rm 1,pAGB}+M_{\rm 2:pAGB}=2.4 M_\odot$,
and examine how much mass it accretes from the evolving tertiary star
(technically we take it as a NS in {\it binary\_c}).
As the tertiary star evolves and fills its Roche lobe, it transfers a part
of its envelope to the inner binary.
To account for the mass loss enhancement from the rotation around the
massive inner binary \citep{ToutEggleton1988}, we take the giant branch
wind multiplier to be $=10$ in {\it binary\_c}.
The tertiary ends this phase of evolution as a He WD of mass
$M_{\rm 3:f}= 0.41M_\odot$; this will be its final mass.

With {\it binary\_c} we find that the tertiary star transfers a mass of
$M_{\rm acc:t} = 0.55 M_\odot$ via RLOF to the inner binary (phase VI in
Fig.\ref{fig:mech}).
Some fraction, $0.05-0.3$, of this mass might be blown away by jets launched
by the accretion discs around the two stars.
For that, we take $0.44 M_\odot$ to be accreted
by the central binary system.
We assume the mass accreted according to the mass of each star;
the secondary, $M_2$, accretes $= 0.18M_\sun$ whereas the ONeMg WD, $M_1$,
accretes $= 0.26M_\sun$ and experiences an AIC to form a NS of mass
$M_{\rm NS:pAIC}\approx 1.5 M_\odot$.
As mentioned in section \ref{sec:scenario}, we find using 
{\it binary\_c} that the duration of the RLOF allows a mass transfer from
the tertiary to the primary at a rate appropriate to cause an AIC
\citep{Taurisetal2013, Ablimit&Li2015}.
The reduction in the mass of $M_1$ is due to energy carried by neutrinos
emitted by the cooling NS.
Considering the mass loss to jets and neutrinos, the orbital separation of
the tertiary star from the center of mass of the inner binary is somewhat
larger than what the {\it binary\_c} simulation gives.
We find the separation at the end of the tertiary evolution and AIC stage
to be $a_{\rm 12,3:pAIC} = 157 R_\odot$ (phase VIII in Fig.\ref{fig:mech}).
The inner binary system is composed now of a NS of mass 
$M_{\rm 1:pAIC}= 1.5 M_\odot$ and a MS star of mass
$M_{\rm 2:pAIC}= 1.18 M_\odot$ orbiting each other with an orbital
separation of $a_{\rm 1,2:pAIC}= 16.5 R_\odot$ .

We next run {\it binary\_c} for the inner system as the secondary star
evolves off the MS and forms a He core
(phase IX in Fig.\ref{fig:mech}).
As discussed in section \ref{sec:scenario}, the mass transfer can
occur in one of three processes: stable RLOF with some mass loss from
the second Lagrangian point, an efficient CE phase, or a GEE.
In the case of a stable RLOF with mass loss from the second Lagrangian point,
we estimated in section \ref{sec:scenario} that to reach the final inner
binary separation of $a_{\rm 1,2:f}\approx 6-8R_\odot$, a mass of
$\sim 0.1 M_\odot$ should be lost from the second Lagrangian point. 
On the other hand, evolving the inner binary using {\it binary\_c}
results in a CE phase.
This CE phase must be very efficient in removing the secondary envelope,
a process we attribute to jets blown by the NS \citep{Papishetal2013}.
Similar to the AGB interaction phase, we can only mimic this high
efficiency by taking a high value of $\alpha_{\rm CE}$.
Using {\it binary\_c} for our case study with $\alpha_{\rm CE}=10$ we find
at the end of this stage the final masses of the NS and the secondary star,
now a He WD, to be $M_{\rm NS:f}= 1.5 M_\odot$, and $M_{\rm 2:f}= 0.2 M_\odot$,
respectively, and their orbital separation to be
$a_{\rm 1,2:f}= 6.3R_\odot$.
Due to the mass loss by the inner binary system, the orbital separation of
the tertiary star increases from $ a_{\rm 12,3:pAIC}=157 R_\odot$ to
$a_{\rm 12,3:f}= 231 R_\odot$.
This is the final, and observed, stage of the system
(phase X in Fig.\ref{fig:mech}; sixth column of table \ref{tab:parameters}).
The masses and orbital separations at the five phases are summarized in
Table \ref{tab:parameters}.

We note that due to some uncertainties in some of the processes, e.g.,
how much mass is lost in the accretion process on to the NS, the final
parameters, e.g., final mass of the NS, are not certain.
Therefore some differences between our derived final parameters and the
observed ones are not significant.
Such is the difference between the observed NS mass and our value in the
case study, $1.438 M_\odot$ and $1.5 M_\odot$, respectively.

Using the {\it binary\_c} code for different post primary AGB
parameters of the triple system, we find that 
although the parameter space is narrow there is flexibility
in setting the post-AGB parameters to achieve the observed
PSR J0337+1715 system.
Namely, there is no need to heavily fine tune the post-AGB
parameters that can span the following ranges:
$M_{\rm 1:pAGB}\approx 1.25-1.4 M_\odot$,
$M_{\rm 2:pAGB}\approx 0.9-1.2 M_\odot$,
$M_{\rm 3:pAGB}\approx 1.3-1.4 M_\odot$,
$a_{\rm 12,3:pAGB}\approx 260-300R_\odot$
and $a_{\rm 1,2:pAGB}\approx 17-21R_\odot$.
We also find that a tertiary post primary-AGB eccentric orbit can be
circularized as the tertiary becomes a giant and tidally interacts
with the inner binary.
For example, taking for the post primary-AGB tertiary orbit an
eccentricity and semimajor axis of $e_{\rm 12,3:pAGB}=0.8$ and
$a_{\rm 1,23:pAGB}\approx 550R_\odot$,
respectively, leads to the desired post-tertiary parameters.
The reason is the strong tidal interaction at
periastron passages during the RGB phase of the tertiary star.

\begin{figure}[h!]
\centering
\includegraphics[width=170mm]{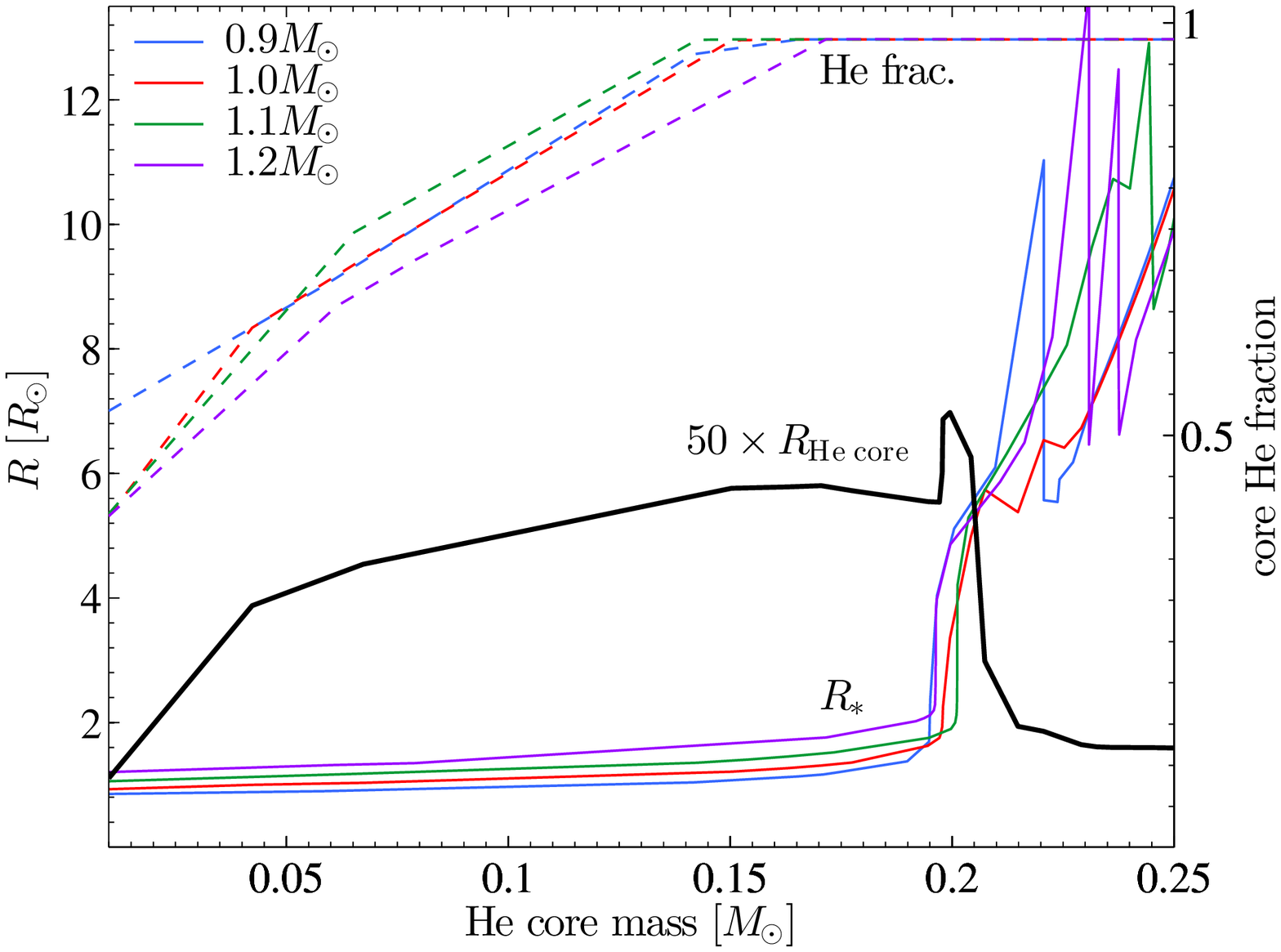}
\caption{The evolution of a $0.9-1.2M_\odot$ star calculated with MESA
(\citealt{Paxton2011}), from zero-age main sequence (ZAMS) until the first
expansion phase where the star has a helium core, as a function of the He
core mass.
The colors are as follows: blue is for a $0.9M_\odot$ star, red is for
a $1.0M_\odot$ star, green is for a $1.1M_\odot$ star and purple is for
a $1.2M_\odot$ star.
The thick solid black line depicts the radius of the helium core for the case of
a $1.0M_\odot$ star.
The left vertical axis relates to the radius of the star (thin solid lines)
and the radius of the helium core (thick solid black line), and the right
vertical axis relates to the He fraction in the core (dashed lines).
These plots underline the well known behaviour of large stellar expansion
in low mass stars of $\approx 1M_\odot$ when the helium core mass is
$\approx 0.19-0.2M_\odot$.
This has great importance for our newly proposed scenario as this behaviour
accounts for the mass of the secondary star in the PSR J0337+1715 system.
}
\label{fig:M2_He}
\end{figure}

The final mass of the secondary star, $M_2=0.197 M_\odot$, does
not depend much on the post primary-AGB separation as well.
During the post MS evolution of the secondary star it substantially expands
as a giant when its helium-core mass is $\approx 0.2M_\odot$
\citep{Refsdal&Weigert1971, Webbinketal1983, Kippenhahn&Weigert1990},
and a strong interaction with the NS companion removes the secondary
envelope (either  in a CE phase or by RLOF, see section \ref{sec:scenario}).
This rapid expansion with little increase in core mass has been used
before to explain the mass of the He WD companion of the MSPs
\citep{Savonije1987, Tauris&Savonije1999, Istrateetal2014}.
\cite{Savonije1987} specifically studied a case of binary pulsars that result
from mass transfer from a (sub)giant to a more massive compact companion.
He found that the degenerate helium core of the (sub)giant determines the
radius and luminosity of the remnant.
We point out that in our model we concentrate on a scenario where the
secondary star had begun evolving after the AIC of the NS primary.
Though we raise the possibility for AIC from the secondary star, we
find the AIC earlier through the tertiary star to be more plausible
(see section \ref{sec:scenario}).
We use the Modules for Experiments in Stellar Astrophysics (MESA),
version 5819 \citep{Paxton2011} to evolve few models that are presented
in Fig. \ref{fig:M2_He}.
This figure emphasizes the, well known but crucial for our proposed
scenario, behaviour described above.
Namely, that stars with an initial mass as the secondary star in our model,
$\approx 0.9-1.2M_\odot$, expand substantially once a helium-core in the mass
range of $\approx 0.19-0.2M_\odot$ is formed.

\section{SUMMARY}
\label{sec:Summary}

We proposed a scenario for the triple system PSR~J0337+1715
composed of a pulsar orbited by two white dwarfs (WDs)
\citep{Kaplanetal2014,Ransometal2014, TaurisHeuvel2014}.
Our proposed scenario is described in Fig. \ref{fig:mech}, and the
values for a case study at the different evolutionary phases are
given in Table \ref{tab:parameters}.

The triple pulsar system PSR J0337+1715 raises two puzzles.
(1) What initial setting can lead to a very close inner WD,
the secondary, orbiting a pulsar, and a WD, the tertiary, at
an orbital separation of about the size of a giant star.
It is clear that the secondary WD could not have evolved through a
regular common envelope (CE) phase with the progenitor of the neutron star
(NS) without the influence of the tertiary star.
(2) The triple system is coplanar, indicating that no violent event could
have taken place, and the system started coplanar.
This suggests that the NS was formed by accretion induced
collapse (AIC) rather than by a core collapse supernova.
We propose that the AIC is caused by mass transfer from the tertiary star
during its red giant phase (phases VI-VIII in Fig. \ref{fig:mech}).
Our scenario can also accommodate an AIC induced by mass transfer from 
the secondary star.
We do not focus on this scenario since we find the AIC at an earlier stage
through mass transfer from the tertiary star to be more plausible.

To solve these puzzles we have introduced to the scenario two new
ingredients. The first one is the entrance of a relatively close
binary system (the tight binary system) into the envelope of the
primary star during its giant phase. Namely, the two lower-mass
stars enter the primary envelope together, rather than one after
the other as in the scenario of \cite{TaurisHeuvel2014}.
The second new ingredient is the process of grazing envelope evolution
(GEE; \citealt{Soker2015}).

We assumed that the secondary-tertiary tight binary system evolves
in a GEE, such that the giant primary envelope outside the center
of mass of the tight binary system is removed.
The spiralling-in process while grazing the giant envelope is a GEE.
Namely, the outer star, or a tight binary system, is evolving in a
`just enters a CE phase'.
Jets launched by the compact companion, or by one or two of the stars
in a tight binary, remove the giant envelope outside the orbit.
If one is to use the $\alpha_{\rm CE}$ parameter, then this efficient
envelope removal increases the effective value of the CE parameter to
values of $\alpha_{\rm CE} > 1$.
In the tight binary system only the star that is closer to
the center of the primary star suffers the full gas dynamical
friction by the envelope. The other star is in a region where the
envelope gas has been removed already.
We consider here an outcome where the tight binary spirals-in and tidally
breaks-up, such that the lighter secondary star in-spirals and
forms a binary system with the massive primary core, whereas the
heavier tertiary star ends on a larger orbit.

We then used the {\it binary\_c} numerical code
\citep{Izzardetal2004, Izzardetal2006, Izzardetal2009} to further
evolve the system, first the tertiary star and then the secondary
star, as described in section \ref{sec:pulsar_formation}.
Here as well we had to assume that tidal spin-up and/or jets launched by
the compact star aid in removing the envelope.
This implies that we can use a value of $\alpha_{\rm CE} > 1$.
We used values of $\alpha_{\rm CE:IV} = 5$ when we evolve the primary star
(IV. AGB interaction in Fig. \ref{fig:mech}) and $\alpha_{\rm CE:IX} = 10$
when we evolved the secondary star (IX. Secondary evolution in Fig. \ref{fig:mech}).
It is quite possible that during the later CE phases, IV and IX in Fig.
\ref{fig:mech}, not only do the jets aid in removing the envelope,
but the jets are very efficient in doing so and lead to a GEE
\citep{Soker2015}, at least during part of the strong interaction phase.
On the other hand we do not rule out that the secondary evolution
(phase IX) occurs through RLOF with mass loss enhancement through the second
Lagrange point due to the corona of the giant secondary star.
Both cases of secondary evolution, either CE or RLOF, will result
in the shrinkage of the inner binary orbital separation.
We also show that no heavy fine-tuning is required to
achieve the triple system PSR~J0337+1715 when one considers tidally
enhanced mass loss and jet-removal of the envelope.
The parameter space to reproduce the system is not large, though.

Our results have more general implications in showing
that a CE with a binary system entering the envelope
of a giant star can lead to a rich variety of evolved triple systems.
The survivability of the three stars in such an evolution can be
accounted for if the removal of an envelope of a giant star with tidal
spin-up and jets is considered.

\textit{Acknowledgements.}
We thank an anonymous referee for detailed comments that substantially
improved both the scientific content and the presentation of our results.
We also thank Erez Michaely, Hagai Perets
and Robert Izzard for very valuable discussions and suggestions.
This research was supported by a generous grant from the president
of the Technion Professor Peretz Lavie, and by the Asher Fund for
Space Research at the Technion.


\label{lastpage}


\begin{thebibliography}{}\addcontentsline{toc}{section}{References}

\bibitem[Ablimit \& Li(2015)]{Ablimit&Li2015} Ablimit, I., \& Li, X.-D.\ 2015, \apj, 800, 98 

\bibitem[Akashi et al.(2015)]{Akashietal2015} Akashi, M., Sabach, E., 
Yogev, O., \& Soker, N.\ 2015, arXiv:1502.05541  

\bibitem[Bhattacharya \& van den Heuvel(1991)]{Bhattacharya&vandenHeuvel1991} Bhattacharya, D., \& van den Heuvel, E.~P.~J.\ 1991, \physrep, 203, 1 

\bibitem[Canal et al.(1980)]{Canaletal1980} Canal, R., Isern, J., 
\& Labay, J.\ 1980, \apjl, 241, L33 

\bibitem[Champion et al.(2008)]{Campionetal2008} Champion, D.~J., Ransom, S.~M., Lazarus, P., et al.\ 2008, Science, 320, 1309

\bibitem[Dessart et al.(2006)]{Dessartetal2006} Dessart, L., Burrows, 
A., Ott, C.~D., et al.\ 2006, \apj, 644, 1063

\bibitem[Freire et al.(2011)]{Freireetal2011} Freire, P.~C.~C., Bassa, C.~G., Wex, N., et al.\ 2011, \mnras, 412, 2763

\bibitem[Hurley et al.(2002)]{Hurleyetal2002} Hurley, J.~R., Tout,
C.~A., \& Pols, O.~R.\ 2002, \mnras, 329, 897

\bibitem[Istrate et al.(2014)]{Istrateetal2014} Istrate, A.~G., Tauris, T.~M., \& Langer, N.\ 2014, \aap, 571, AA45 

\bibitem[Ivanova et al.(2013)]{Ivanovaetal2013} Ivanova, N., Justham, S., Chen, X., et al.\ 2013, \aapr, 21, 59

\bibitem[Izzard et al.(2004)]{Izzardetal2004} Izzard, R.~G., Tout,
C.~A., Karakas, A.~I., \& Pols, O.~R.\ 2004, \mnras, 350, 407

\bibitem[Izzard et al.(2006)]{Izzardetal2006} Izzard, R.~G., Dray, L.~M., Karakas, A.~I., Lugaro, M., \& Tout, C.~A.\ 2006, \aap, 460, 565

\bibitem[Izzard et al.(2009)]{Izzardetal2009} Izzard, R.~G., Glebbeek, E., Stancliffe, R.~J., \& Pols, O.~R.\ 2009, \aap, 508, 1359

\bibitem[Kaplan et al.(2014)]{Kaplanetal2014} Kaplan, D.~L., van
Kerkwijk, M.~H., Koester, D., et al.\ 2014, \apjl, 783, L23

\bibitem[Kippenhahn \& Weigert(1990)]{Kippenhahn&Weigert1990} Kippenhahn, R., \& Weigert, A.\ 1990, Stellar Structure and Evolution, XVI, 468 pp.~192 figs..~ Springer-Verlag Berlin Heidelberg New York.~Also Astronomy and Astrophysics Library, 

\bibitem[Kitaura et al.(2006)]{Kitauraetal2006} Kitaura, F.~S., Janka, H.-T., \& Hillebrandt, W.\ 2006, \aap, 450, 345

\bibitem[Livio et al.(1979)]{Livioetal1979} Livio, M., Salzman, J., 
\& Shaviv, G.\ 1979, \mnras, 188, 1 

\bibitem[Miller et al.(2005)]{Milleretal2005} Miller, M.~C., Freitag,
M., Hamilton, D.~P., \& Lauburg, V.~M.\ 2005, \apjl, 631, L117

\bibitem[Nomoto \& Kondo(1991)]{Nomoto&Kondo1991} Nomoto, K., \& Kondo, Y.\ 1991, \apjl, 367, L19

\bibitem[Nomoto et al.(1979)]{Nomotoetal1979} Nomoto, K., Miyaji, S., 
Sugimoto, D., \& Yokoi, K.\ 1979, IAU Colloq.~53: White Dwarfs and Variable Degenerate Stars, 56

\bibitem[Papish et al.(2013)]{Papishetal2013} Papish, O., Soker, N., 
\& Bukay, I.\ 2013, arXiv:1309.3925

\bibitem[Paxton et al.(2011)]{Paxton2011} Paxton, B., Bildsten,
L., Dotter, A., et al.\ 2011, \apjs, 192, 3

\bibitem[Pijloo et al.(2012)]{Pijlooetal2012} Pijloo, J.~T., Caputo, D.~P., \& Portegies Zwart, S.~F.\ 2012, \mnras, 424, 2914

\bibitem[Portegies Zwart et al.(2011)]{PortegiesZwartetal2011} Portegies Zwart, S., van den Heuvel, E.~P.~J., van Leeuwen, J.,
\& Nelemans, G.\ 2011, \apj, 734, 55

\bibitem[Ransom et al.(2014)]{Ransometal2014} Ransom, S.~M., Stairs, I.~H., Archibald, A.~M., et al.\ 2014, \nat, 505, 520

\bibitem[Rappaport et al.(2013)]{Rappaportetal2013} Rappaport, S., Deck, K., Levine, A., et al.\ 2013, \apj, 768, 33

\bibitem[Refsdal \& Weigert(1971)]{Refsdal&Weigert1971} Refsdal, S., \& Weigert, A.\ 1971, \aap, 13, 367

\bibitem[Savonije(1987)]{Savonije1987} Savonije, G.~J.\ 1987, \nat, 
325, 416

\bibitem[Sesana et al.(2009)]{Sesanaetal2009} Sesana, A., Madau, P.,
\& Haardt, F.\ 2009, \mnras, 392, L31

\bibitem[Soker(2014)]{Soker2014} Soker, N.\ 2014, arXiv:1404.5234

\bibitem[Soker(2015)]{Soker2015} Soker, N.\ 2015, \apj, 800, 114 

\bibitem[Tauris(2015)]{Tauris2015} Tauris, T.~M.\ 2015, arXiv:1501.03882

\bibitem[Tauris \& Savonije(1999)]{Tauris&Savonije1999} Tauris, T.~M., \& Savonije, G.~J.\ 1999, \aap, 350, 928

\bibitem[Tauris  \& van den Heuvel(2006)]{Tauris&vandenHeuvel2006} Tauris, T.~M., \& van den Heuvel, E.~P.~J.\ 2006, Compact stellar X-ray sources, 623

\bibitem[Tauris \& van den Heuvel(2014)]{TaurisHeuvel2014} Tauris, T.~M., \& van den Heuvel, E.~P.~J.\ 2014, \apjl, 781, L13

\bibitem[Tauris et al.(2013)]{Taurisetal2013} Tauris, T.~M., Sanyal, D., Yoon, S.-C., \& Langer, N.\ 2013, \aap, 558, AA39

\bibitem[Tout \& Eggleton(1988)]{ToutEggleton1988} Tout, C.~A., \& Eggleton, P.~P.\ 1988, \mnras, 231, 823

\bibitem[Webbink et al.(1983)]{Webbinketal1983} Webbink, R.~F., 
Rappaport, S., \& Savonije, G.~J.\ 1983, \apj, 270, 678









\end{thebibliography}
\end{document}